\documentclass[a4paper]{VisionStyle}
\usepackage{epsfig}

\def\spose#1{\hbox to 0pt{#1\hss}}
\def\lax{$\mathrel{\spose{\lower 3pt\hbox{$\mathchar"218$}}
     \raise 2.0pt\hbox{$\mathchar"13C$}}$}
\def\gax{$\mathrel{\spose{\lower 3pt\hbox{$\mathchar"218$}}
     \raise 2.0pt\hbox{$\mathchar"13E$}}$}
\newcommand{\chandra}{{\it Chandra}}

\newcommand{\rosat}{{\it ROSAT}}
\newcommand{\xmm}{{\it XMM-Newton}}
\newcommand{\einstein}{{\it Einstein}}
\newcommand{\lum}{\thinspace\hbox{$\hbox{erg}\thinspace\hbox{s}^{-1}$}}

\begin{document}

\title{Two Years of the X-ray Sky in M31}

\author{A.K.H. Kong \and M.R. Garcia \and 
  F.A. Primini  \and R. Di\,Stefano \and S.S. Murray} 

\institute{Harvard-Smithsonian Center for Astrophysics, 60 Garden
Street, Cambridge, MA 02138, U.S.A.}

\maketitle 

\begin{abstract}

We have been monitoring M31 with HRC and ACIS onboard \chandra\ regularly in the past
two years. By combining eight \chandra\ ACIS-I observations taken between
1999 to 2001, we have identified 204 X-ray sources within the central
$\sim 17\arcmin\times17\arcmin$ region of M31, with a detection limit of $\sim
1.6\times10^{35}$\lum. Of these 204 sources, 21 are identified with
globular clusters, 2 with supernova remnants (one of them is spatially
resolved with \chandra), and 8 with planetary
nebula. By comparing individual images, about
50\% of the sources are variable in time scales of months. We also
found 14 transients. Combining all the available transients found in
literatures, there are 25 transients in M31 and M32 detected by
\chandra\ and \xmm; we present some of the long-term lightcurves by
using the HRC, ACIS and \xmm\ data. The spectral
shape of 12 sources is shown to be variable,
suggesting that they went through state changes. 
The luminosity
function of all the point sources is consistent with previous
observations (a broken power-law with a luminosity break at
$1.7\times10^{37}$\lum). However, when the X-ray sources in different
regions are
considered separately, different luminosity functions are obtained. This
indicates that the star formation history might be different in different regions.

\end{abstract}

\section{Introduction}
  
\object{M31} was observed by \chandra\ and \xmm\ soon after these
observatories were launched. In
the first observation (8.8 ks) of the core region by \chandra\ in 1999
October, 121 point sources were identified within the central
$17'\times17'$ region and the nucleus was nicely
resolved into five point sources which were not seen in previous
missions (\cite{akong-E3:gar00a}). Moreover, a bright transient was
discovered $\sim 26\arcsec$ from the nucleus. A relatively deeper \xmm\
observation (34.8 ks) was made in 2000 June; 116 sources were detected
down to a limiting luminosity of $6\times 10^{35}$ \lum\ (0.3--12 kpc;
\cite{akong-E3:shi01}) and a pulsating supersoft transient with a
periodicity of $\sim 865$ s was discovered
(\cite{akong-E3:osb01}). Moreover, both \chandra\
(\cite{akong-E3:gar01a}; \cite{akong-E3:pri00}) and \xmm\
(\cite{akong-E3:shi01}) observations confirmed that
the unresolved X-ray emission in the core region is much softer than
most of the resolved X-ray sources in that region. Fifteen X-ray
point sources are newly associated with globular clusters
(\cite{akong-E3:di02}), which when combined with the previous \rosat\ results
(\cite{akong-E3:sup01}) brings the total number of M31 globular clusters with
detected X-ray emission to 48. \chandra\ and \xmm\ also discovered
several bright ($L_X > 10^{37}$ \lum) transients in M31 and M32
(\cite{akong-E3:gar00a}; \cite{akong-E3:gar00b};
\cite{akong-E3:osb01}; \cite{akong-E3:shi01}; \cite{akong-E3:kong01};
\cite{akong-E3:gar01b}). The brightest of these reached a peak
luminosity of $L_X \sim 3\times10^{38}$ \lum (\cite{akong-E3:kong01}).

We report herein a brief summary of our \chandra\ monitoring program
of M31 in the past two years; it will mainly focus on the properties
(identifications, temporal and spectral variability, and luminosity function)
of point sources in the central
$\sim 17'\times 17'$ region of M31 as deduced from eight separate
$\sim 5$ks ACIS-I observations spanning $\sim 1.5$~years. In addition,
we will also present early results of an X-ray resolved supernova
remnant and an updated ``X-ray movie'' of M31 up to 2002 January.

\section{Observations}

\begin{figure}
\begin{center}
\psfig{file=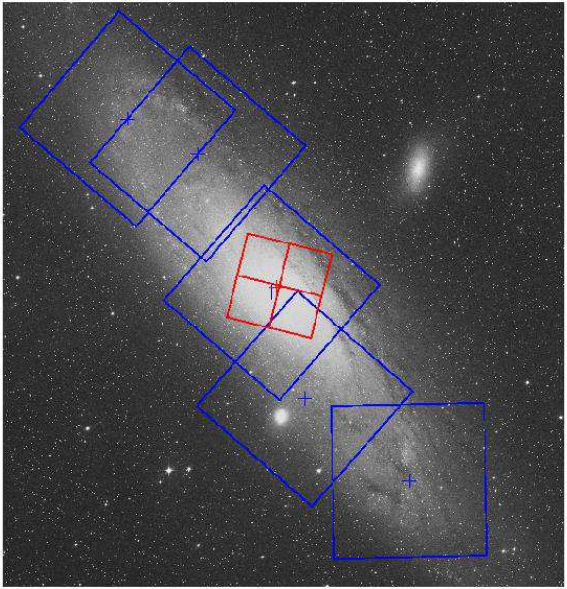,width=8.5cm}
\end{center}
\caption{The field-of-view of HRC-I (blue) and ACIS-I (red) pointings overlaid on
an optical Digitized Sky Survey image of M31. The aim-points are
marked with a cross.}
\label{akong-E3_fig:fig1}
\end{figure}

M31 was observed with \chandra\ regularly as part of the AO-1 and AO-2
GTO program during 1999--2001 (This program continues as GO
observations in AO-3). The program is originally designed to
search for transients in M31. These observations consist of a series
of HRC snapshots ($\sim 1$ ks) which cover the whole galaxy (see
Figure~\ref{akong-E3_fig:fig1}); once a
transient is found, a follow-up ACIS ($\sim 5$ ks) observation then images the
newly-discovered transient, otherwise, ACIS points to the nuclear
region. In this paper, we mainly focus on the ACIS-I (I0, I1, I2 and I3)
data centered in the central $16.9\arcmin\times 16.9\arcmin$ region of M31; this
consists of 8 separate observations from 1999 October to 2001 June,
with exposure times ranging from
4 to 8.8 ks. The actual region covered by the \chandra\ observations is
slightly larger than $16.9\arcmin\times 16.9\arcmin$ due to different roll
angles. Sources near the outer edge of ACIS are therefore not
observable in all eight exposures. In addition, we also make use of
the HRC data and other ACIS pointings to construct the long-term
lightcurves of sources.

\begin{figure*}[!ht]
\begin{center}
\epsfig{file=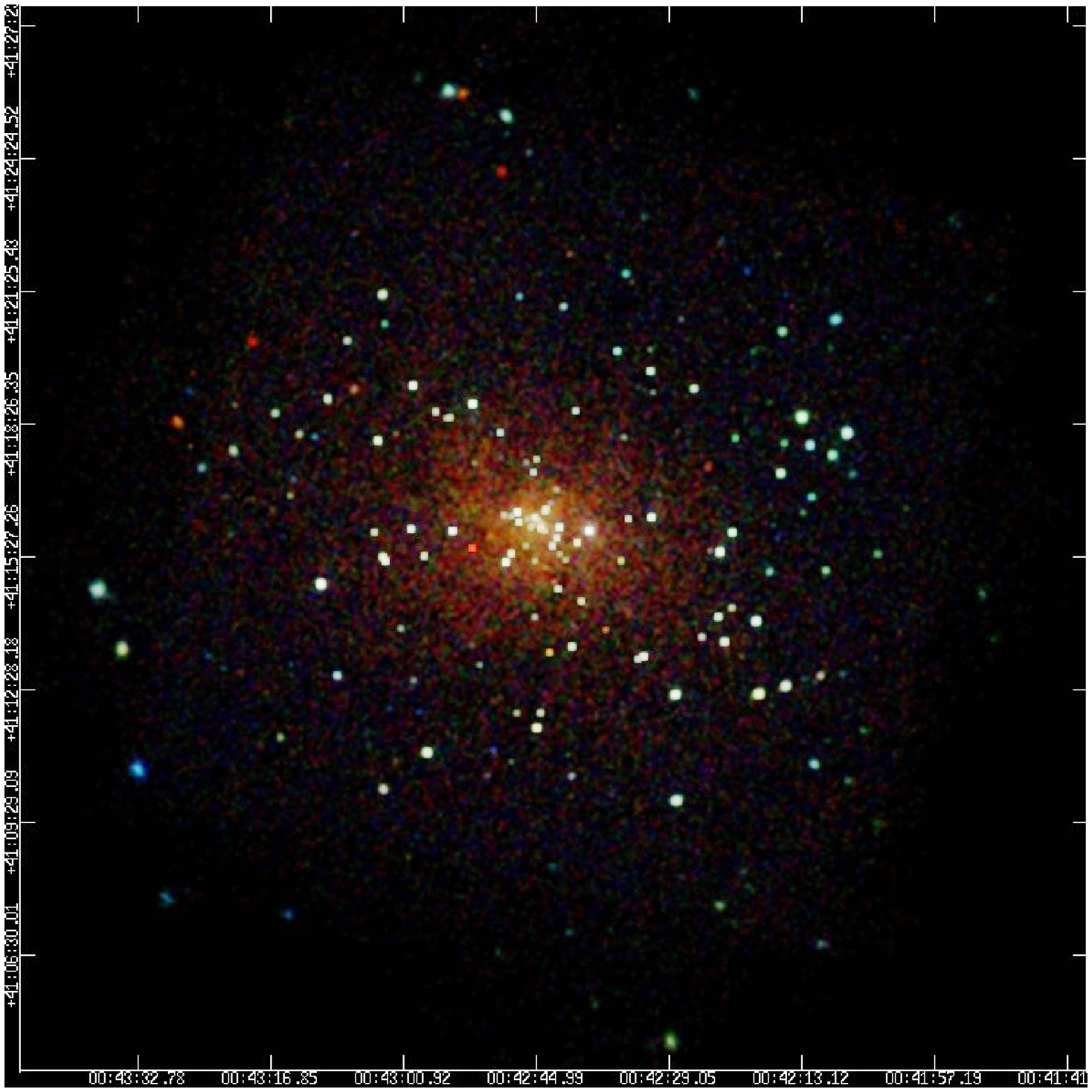,width=13cm}
\epsfig{file=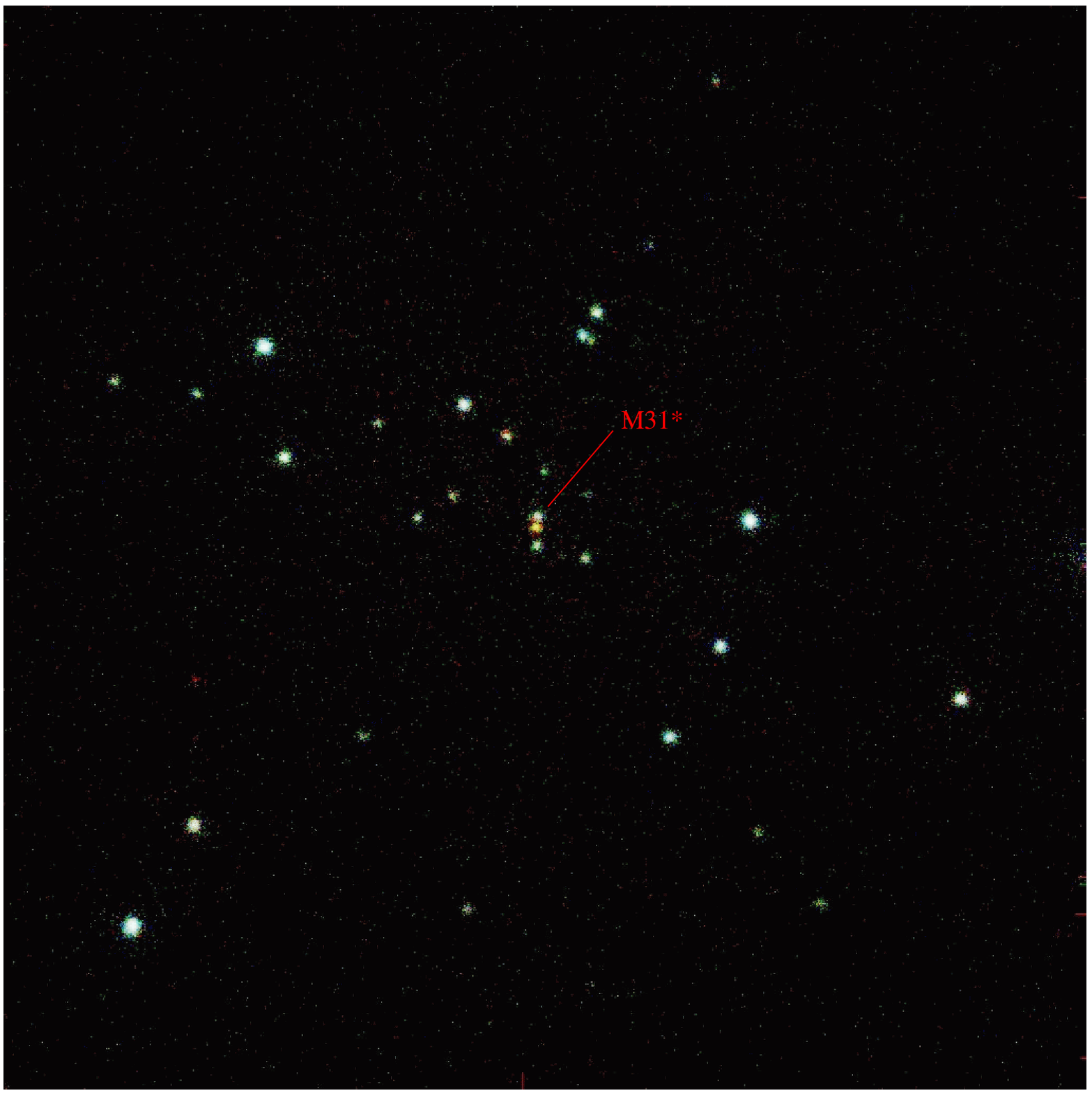,width=8cm}
\end{center}
\caption{Top: Stacked ``true color'' \chandra\ ACIS-I image (39.7 ks) of the central
$\sim 17\arcmin\times17\arcmin$ region of M31. This image was constructed from the
soft (red; 0.3--1 keV), medium (green; 1--2 keV) and hard (blue; 2--7
keV) energy bands. The pixel size is $1.96\arcsec$ and the image has been
smoothed with a Gaussian ($\sigma=0.5\arcsec$) function. Bottom:
Stacked ``true color'' \chandra\ ACIS-I image of the central
$\sim 2\arcmin\times2\arcmin$ region of M31. The pixel size is
$0.123''$. M31$^*$ candidate is marked.}
\label{akong-E3_fig:fig2}
\end{figure*}

In order to create a deep image suitable for the detection of faint
sources, the eight observations were combined into a single stacked
image with a total integration time of 39.7
ks. Figure~\ref{akong-E3_fig:fig2} shows the stacked ``true color''
image of M31, which is a
composite of images from soft (0.3--1.0 keV), middle (1--2 keV) and hard
(2--7 keV) bands. Soft sources appear red, moderately hard sources
appear green, and the hardest sources appear blue. The image has been
corrected for exposure and smoothed slightly with a Gaussian ($\sigma=0.5\arcsec$)
in order to improve the appearance of point sources; diffuse
emission around the nucleus is also clearly shown.  Also shown in
Figure~\ref{akong-E3_fig:fig2} is
the ``true color'' image of the central $2\arcmin\times2\arcmin$ region of M31
with $1/8\arcsec$ pixel resolution, with the possible nuclear counterpart
(M31$^*$) marked (Garcia 2001).

\section{Source Detection and Identification}

Discrete sources in the stacked image were found with WAVDETECT
(\cite{akong-E3:free02}), a wavelet detection algorithm implemented
within CIAO. The central $2\arcmin\times2\arcmin$ region was treated separately by
using the $1/8\arcsec$ image. A total of 204
sources were detected above $2.5\sigma$. 

We cross-correlated the source list with
existing catalogs of M31 objects. We find that 77 \chandra\ sources
have counterparts in the \rosat\ HRI catalog
(\cite{akong-E3:pri93}). The remaining 127 ($=204-77$) \chandra\
sources were not detected in the \rosat\ HRI
catalog, presumably because they are below the \rosat\ detection limit or
are variable.  This \chandra\ catalog extends $\sim 5\times$ fainter
than the \rosat\ HRI catalog. We identify 22 \chandra\ sources with
globular clusters; twelve of these globular clusters are identified as
X-ray sources for the first time. We find two matches with supernova
remnant (SNR). In particular, one of the them is spatially resolved by
\chandra\ (see \S\,5). Eight planetary nebula (PN) are found to be
associated with our \chandra\ sources. However, the X-ray luminosity
of our PN candidates are at least 3 order of magnitude brighter than that in
our Galaxy (e.g. \cite{akong-E3:kas01}); the X-ray colors are generally harder as well. Therefore,
it is possible that these PN are either very unusual or something
other than PN. We suggest that these may be objects similar to
\object{GX\,1+4} in our Galaxy (i.e. symbiotic stars with a neutron star
companion). We searched for matches between stellar nova as listed in the IAUC
contemporaneous to and within the field-of-view of our \chandra\ observations,
and found no matches within $3''$. We also searched for matches with
OB associations, as O and B stars may be moderate X-ray sources ($L_X
< 10^{33}$erg~s$^{-1}$, \cite{akong-E3:ber97}).  While this is well
below our detection limit, a group of O and/or B stars may reach our
detection threshold,  and star forming regions could conceivably
harbor massive X-ray binaries. However, we found no matches within our
search radius of $3''$. Five possible foreground objects are also
found in our field based on their X-ray colors and coincidence of
stars.

\section{Temporal and Spectral Variability}

The eight \chandra\ ACIS-I observations described herein span nearly 2
years from 1999--2001. This is substantially longer than previous
surveys by \rosat\ (2 observations separated by $\sim 1$ year).  In
order to study long-term X-ray variability, we computed a variability
parameter following \cite*{akong-E3:pri93}. We found 99 X-ray variables, corresponding
to $\sim 50$\% of the total. By comparing \rosat\ observations to
\einstein\ observaitions 10~years earlier, \cite*{akong-E3:pri93} found that $\sim
42$\% of the X-ray sources within central $7.5'$ region were variable.
By comparing two \xmm\ observaitions separated by six months
\cite*{akong-E3:osb01} found  that $> 15$\% of the sources in the central $30'$ were
variable. 

\begin{figure*}[!ht]
\begin{center}
\psfig{file=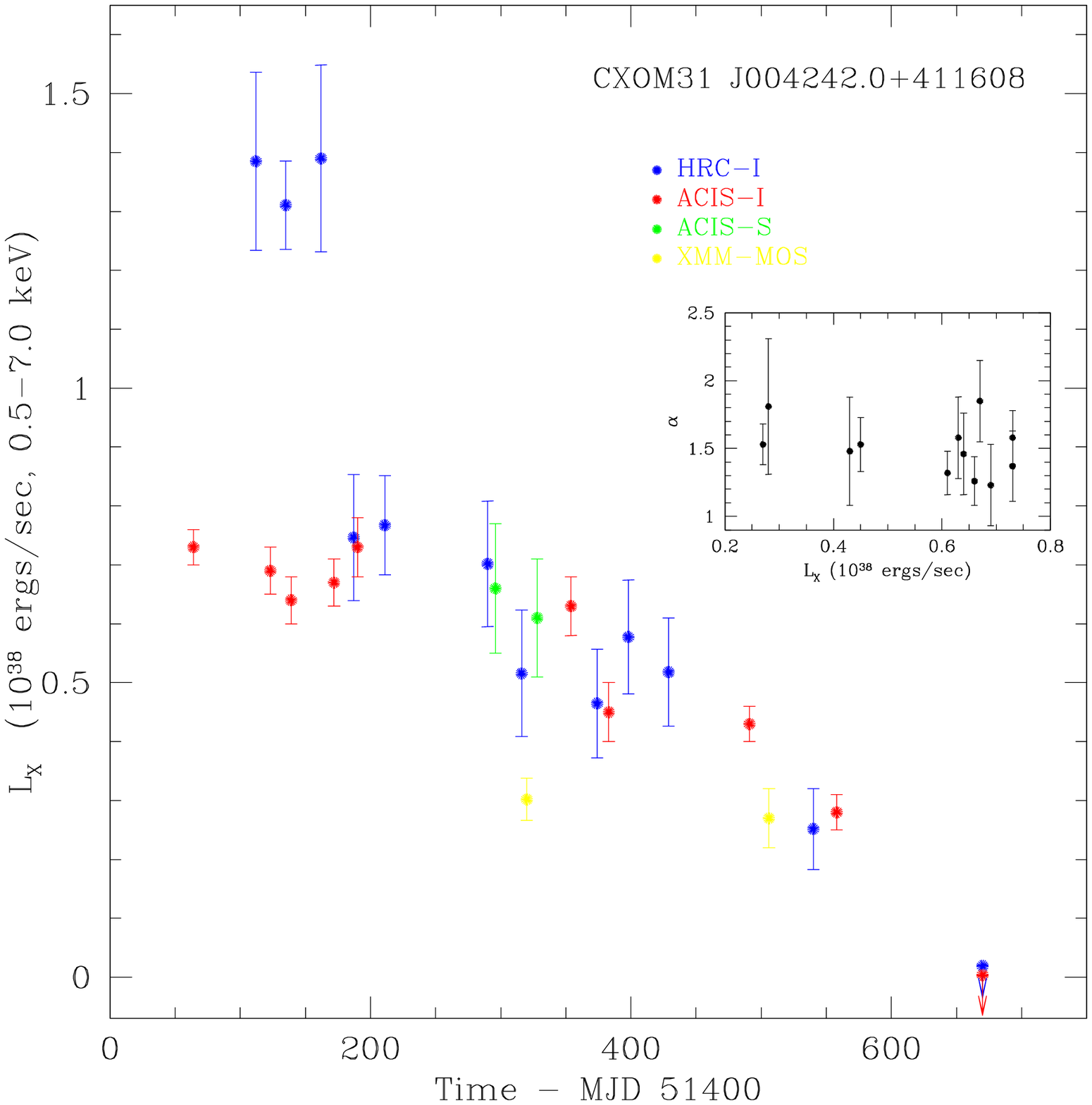,height=7.1cm,width=7.5cm}
\psfig{file=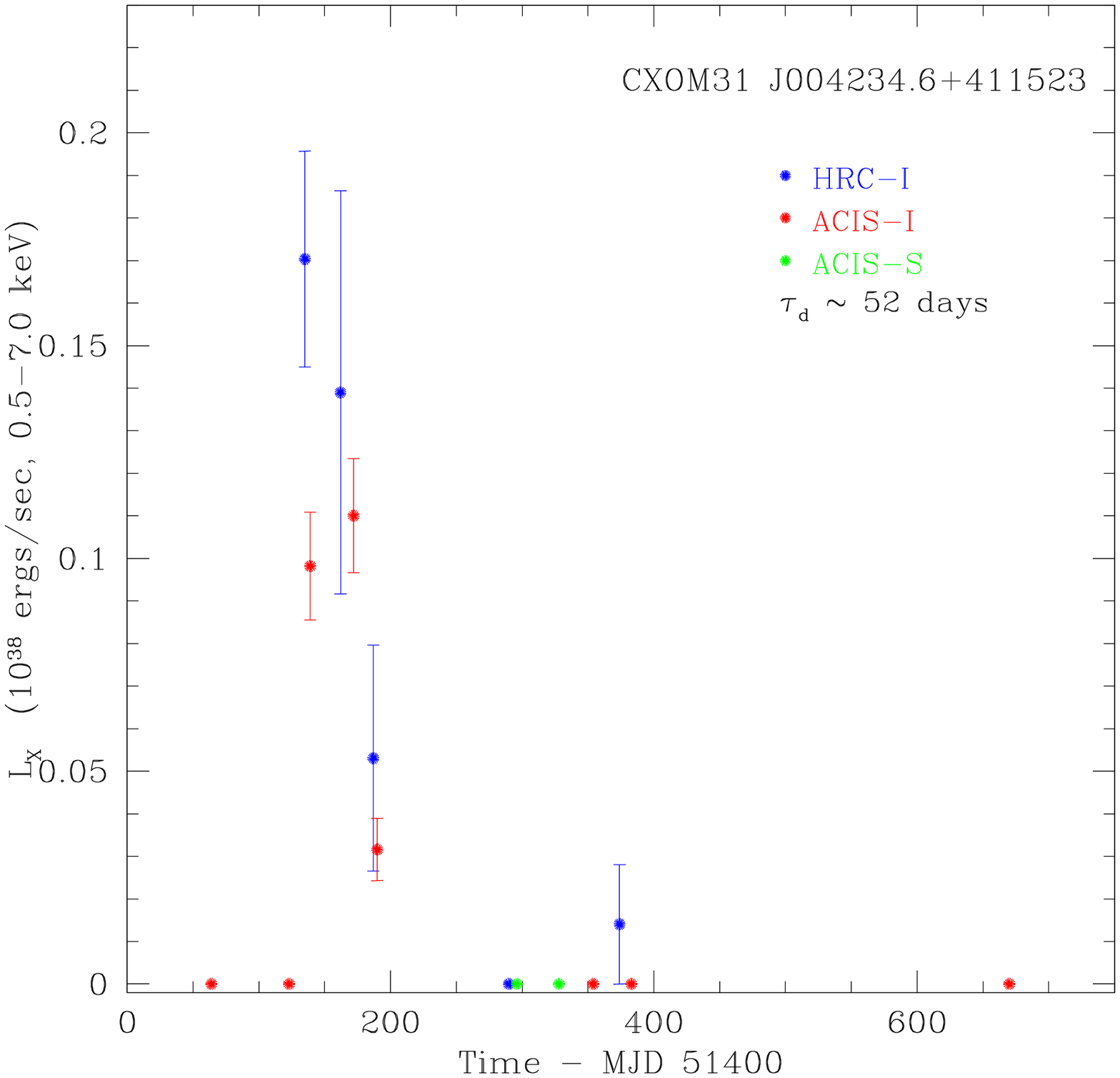,height=7.1cm,width=7.5cm}
\hspace*{0.35cm}\psfig{file=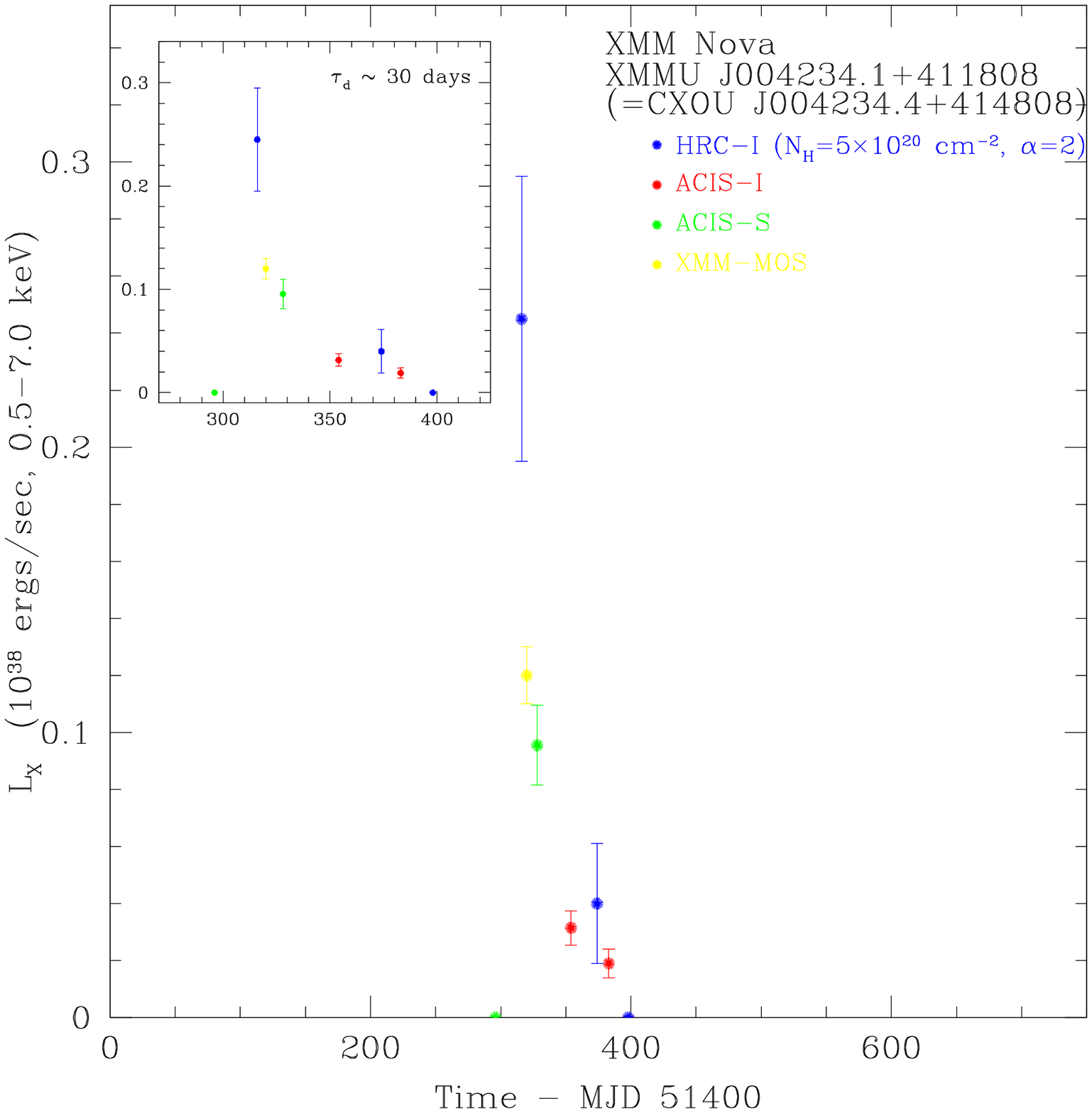,height=7.1cm,width=7.5cm}
\psfig{file=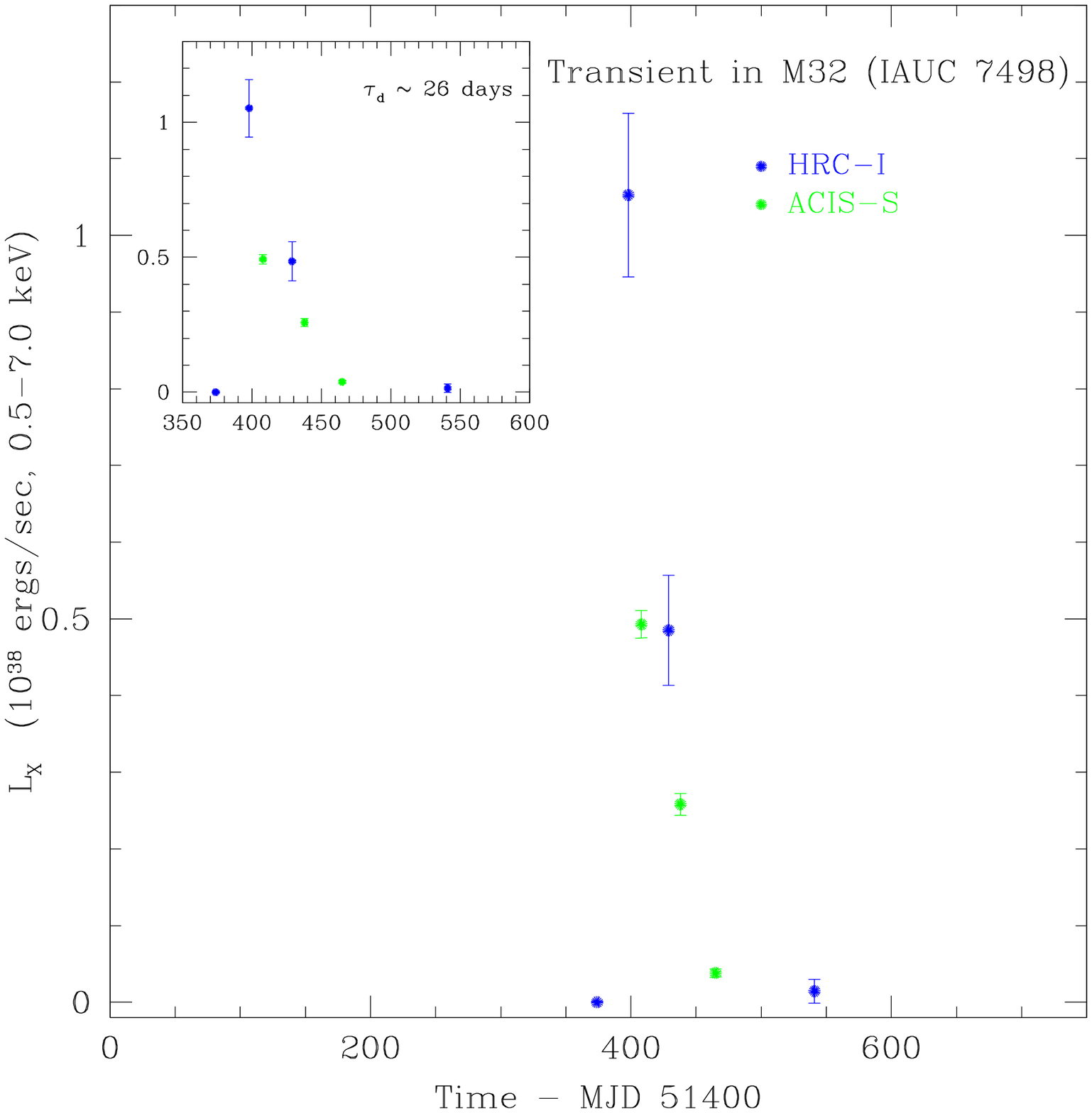,height=7.1cm,width=7.7cm}
\psfig{file=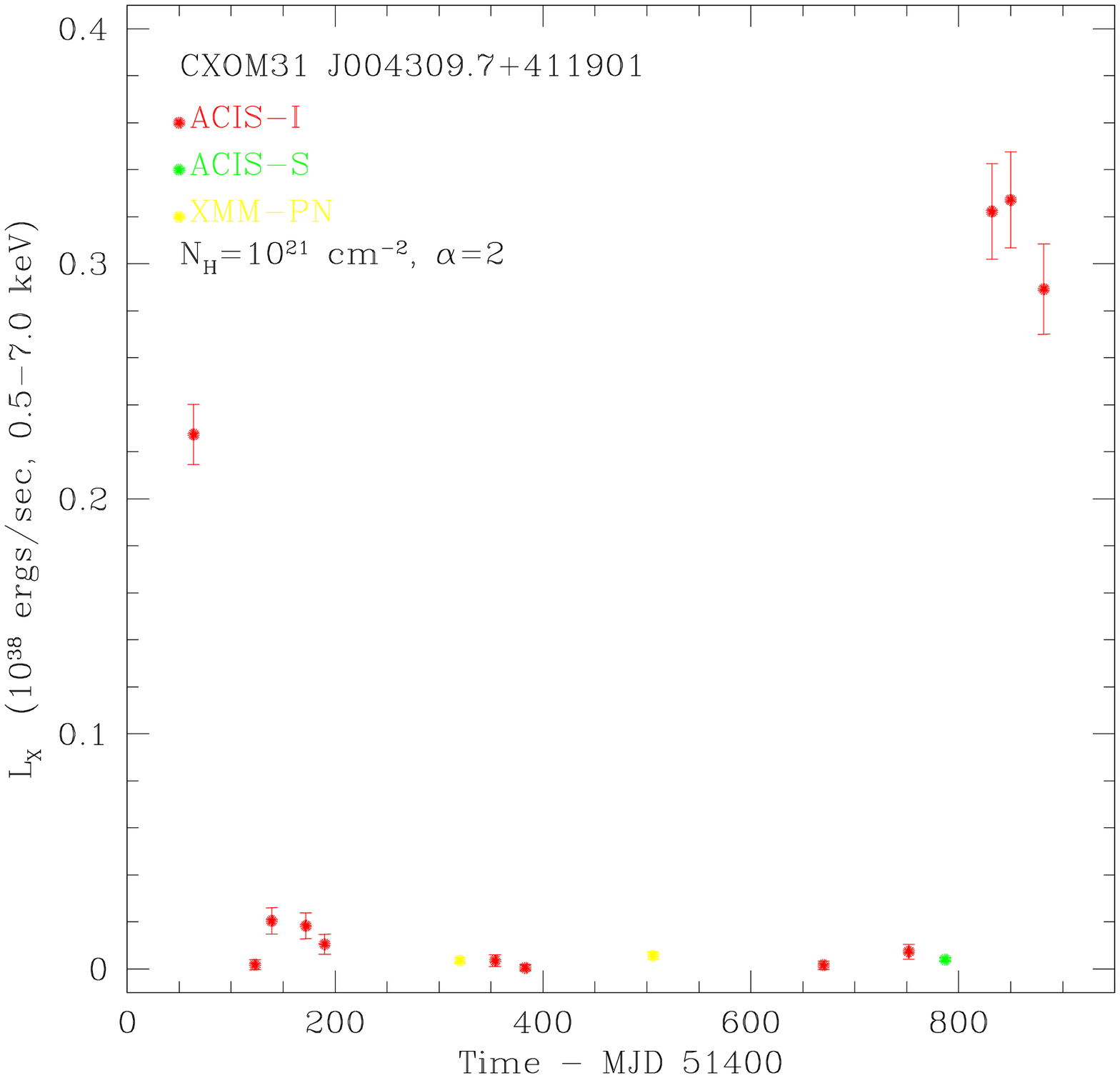,height=7.1cm,width=7.5cm}
\psfig{file=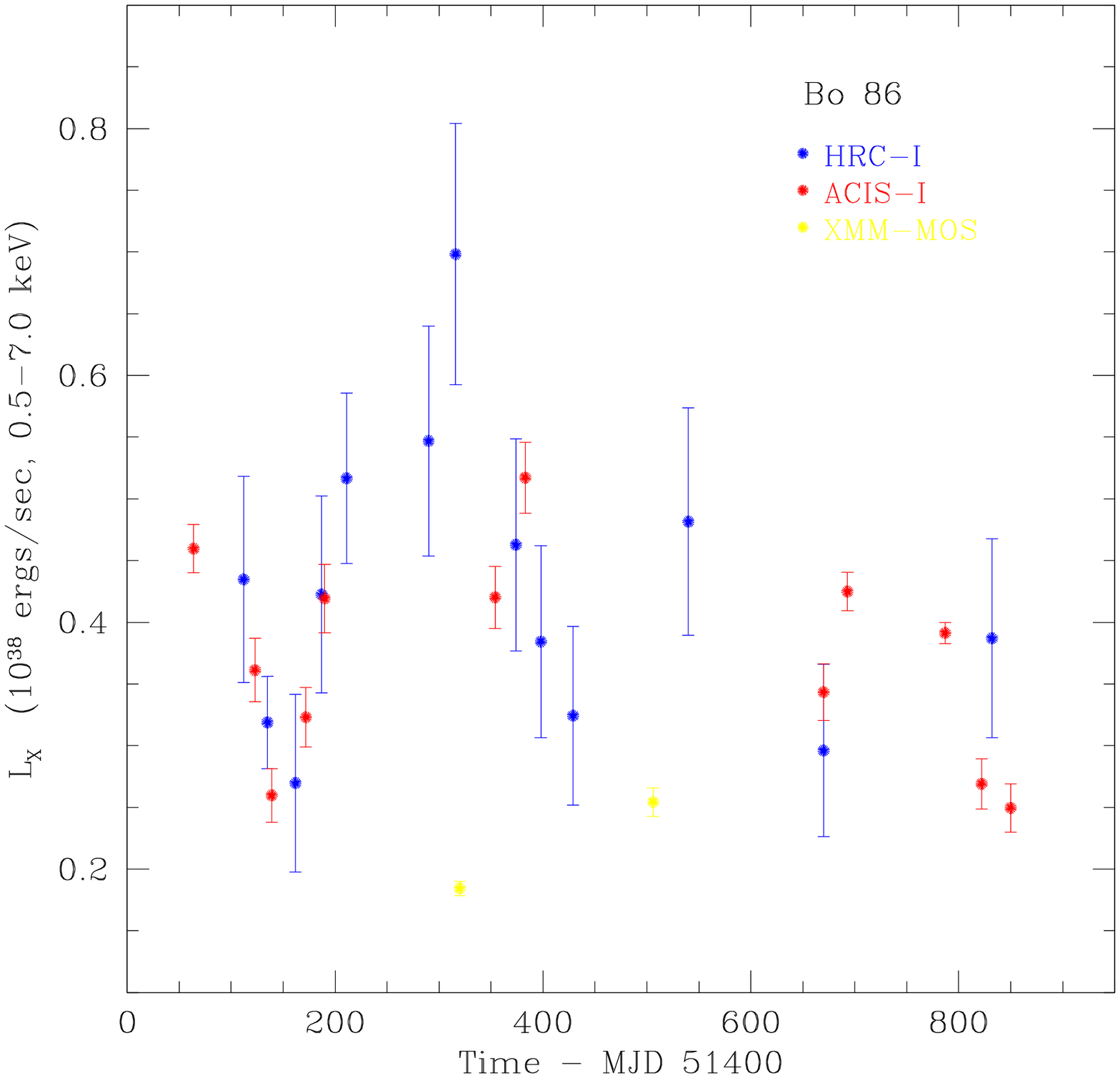,height=7.1cm,width=7.5cm}
\end{center}
\caption{Lightcurves of 5 bright transients and globular cluster
Bo\,86 as seen in the past two years with \chandra\ and \xmm. Observations from
HRC-I (blue), ACIS-I (red), ACIS-S (green) and XMM-MOS/PN (yellow) are included.}
\end{figure*}

\begin{figure*}[!ht]
\begin{center}
\psfig{file=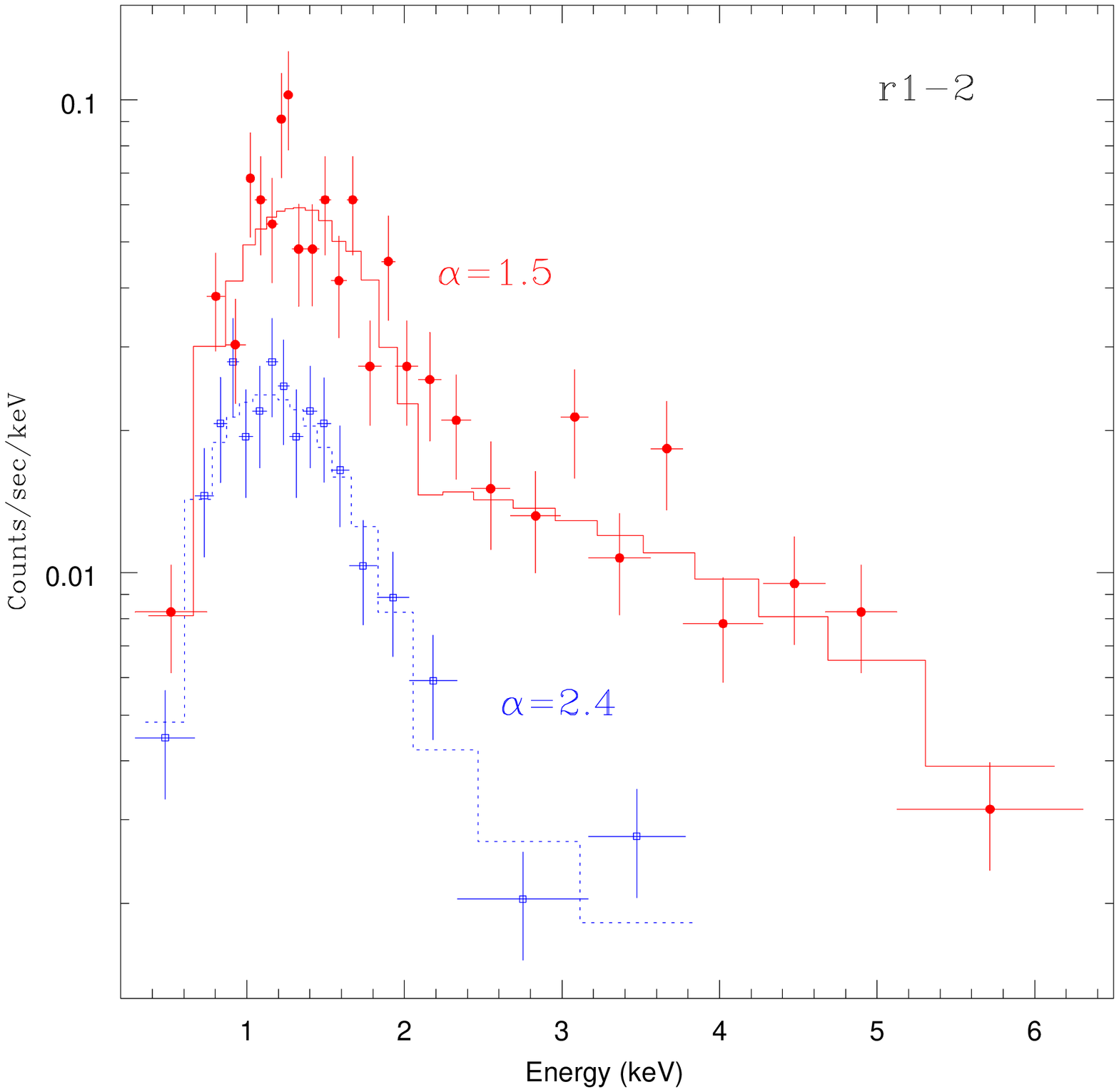,height=6cm,width=5.7cm}
\psfig{file=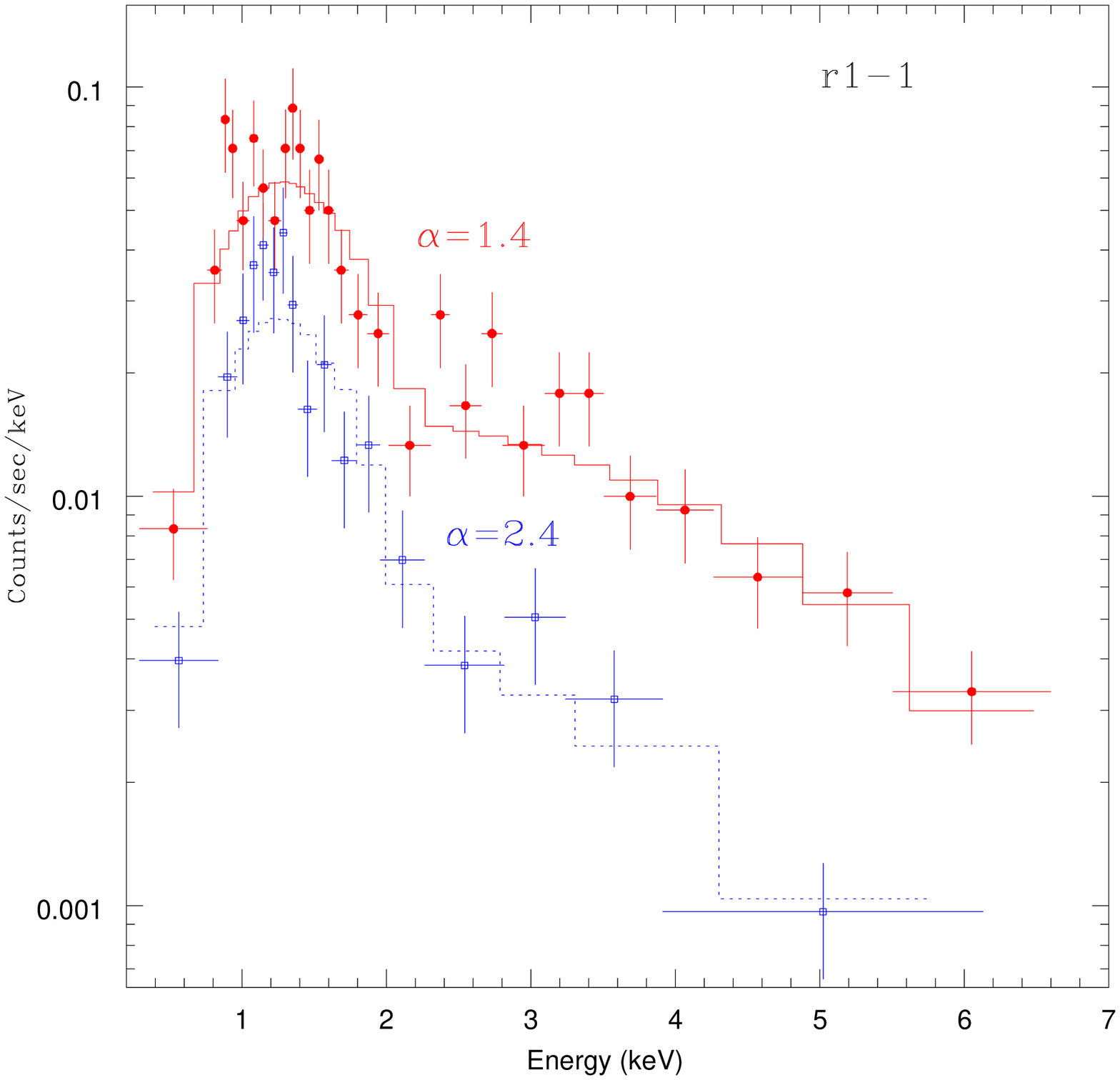,height=6cm,width=5.7cm}
\psfig{file=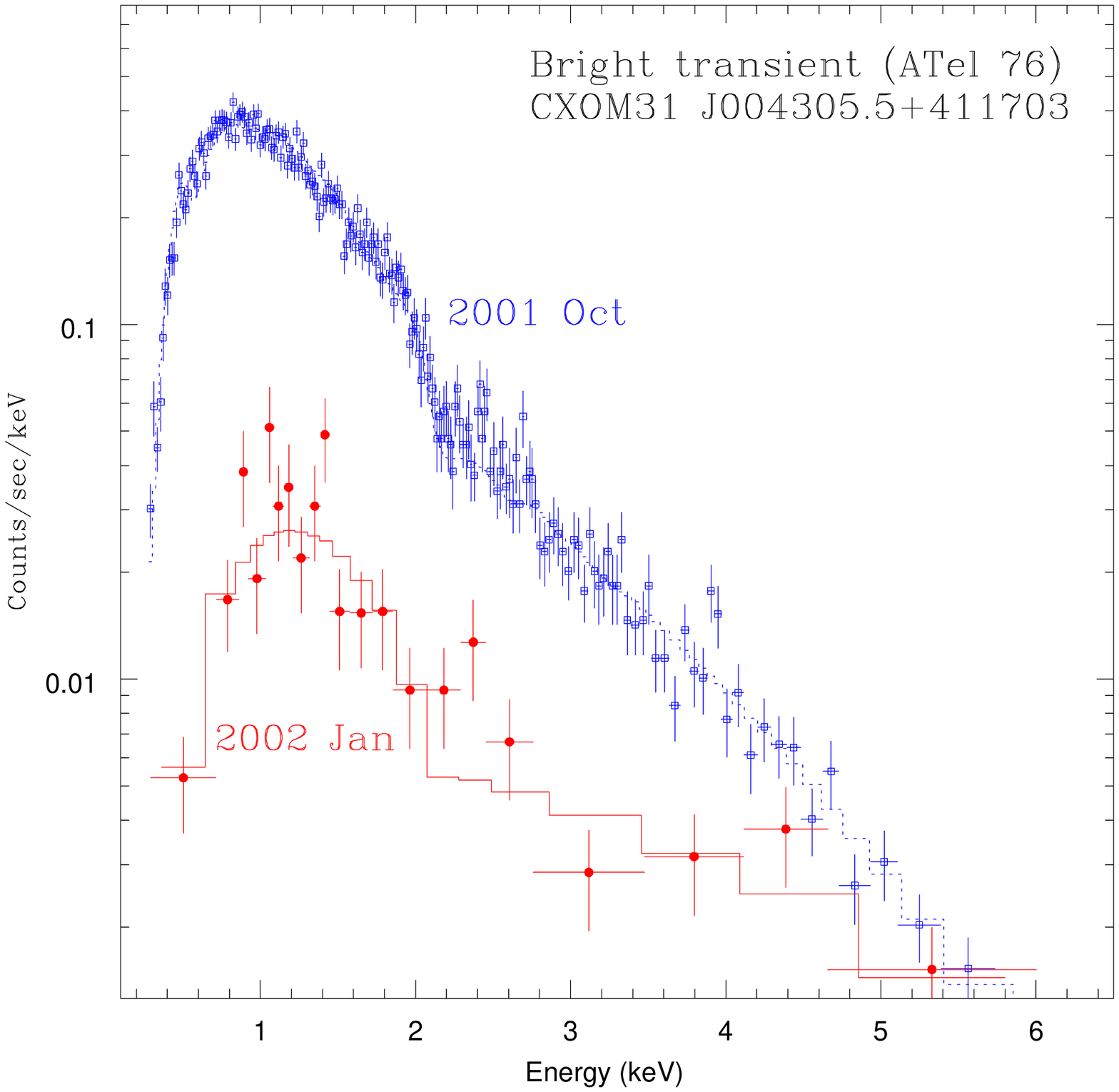,height=6cm,width=5.7cm}
\end{center}
\caption{Spectral changes of three of the sources. The first two from
the left change from lower luminosity at softer state to higher
luminosity with harder state. The bright transient in the right shows a harder
spectrum during its decay from the peak of the outburst.}
\end{figure*}

We have also discovered 13 bright transients for
which the source is
found in at least one observation with luminosity of \gax$\,
5\times10^{36}$ erg s$^{-1}$ and is not detected in at least one of
the other observations. Note that the
luminosity limit covers typical outburst luminosities of soft X-ray
transients and Be/X-ray binaries in our Galaxy. One important
transient was missed by this analysis because it had a
peak luminosity below our ``bright transient'' threshold during the
eight ACIS-I observations considered here.  This
object is XMMU J004234.1+411808 (= CXOM31 J004234.3+411809),
which \cite*{akong-E3:osb01} suggest is an X-ray nova. By adding all
the published X-ray transients in M31 found by \chandra\ and
\xmm\ (\cite{akong-E3:osb01}; \cite{akong-E3:shi01b};
\cite{akong-E3:kong01}; \cite{akong-E3:gar01b};
\cite{akong-E3:tru02}), we conclude that there are 25 ``bright''
transients in M31 (and M32) in the past two years.  

Figure~3 shows five representative lightcurves of transients found in M31
and M32.
The transient discovered with
\chandra\ in the first observation (\cite{akong-E3:gar00a}) remained in
outburst for more than one year and finally turned off in 2001 June
(Figure 3). In addition, three HRC-I data points indicate that the
source underwent flarings or state transitions during the early stage
of the outburst. During the whole outburst, the energy spectrum is
consistent with a power-law with slope of $\sim 1.5$ (see inset of
Figure 3). The remaining three transients show typical lightcurves
with a fast rise followed by an exponential decay like those seen in our
Galaxy (\cite{akong-E3:chen97}). The characteristic decay time is
about 30--50 days. CXOM31
J004309.7+411901 is a recurrent transient; the
source was detected with \einstein, \rosat\ HRI and \rosat\ PSPC at
luminosities ranging from 1 to 8 $\times10^{37}$
\lum. \cite*{akong-E3:sup01} noted variability between PSPC exposures
of a factor of 5. Except for the first \chandra\ observations, the
source underwent a state transition to the low state at $L_X <
10^{36}$ \lum; it re-brightened to $3\times10^{37}$ \lum in 2001 November
and shows decay in 2002 January
observations.  Beginning in the summer of 2001, we also monitored newly
discovered X-ray transients with {\it HST} to search for UV
counterparts. This program allows us to follow the evolution of six
transients in M31 during their outbursts in both X-rays and UV; it
will be the first step in extending the highly successful {\it
RXTE}/ASM and optical follow-up of Galactic X-ray transients to our
nearest neighbor spiral galaxy.  

In addition to transients, we also found several sources which have
interesting long-term lightcurve. For example,
CXOM31\,J004218.5+411758 is in the M31 globular cluster Bo\,86 and shows
a possible $\sim 200$ day modulation.  This is reminiscent of the Galactic source
\object{4U\,1820--30} in the globular cluster \object{NGC~6624}, which
has a 176-d long-term modulation. It is worth noting that  X-ray variability on long
time-scales (from days to years) has been found in many low-mass and
high-mass X-ray binaries in our Galaxy (see e.g. \cite{akong-E3:kong00}).

\begin{figure*}[!ht]
\begin{center}
\epsfig{file=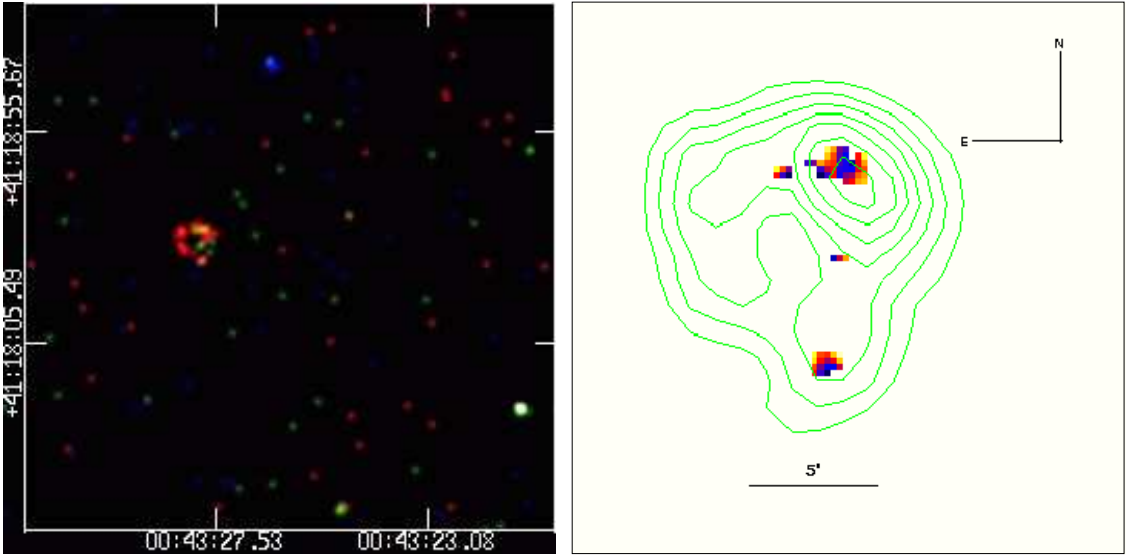,width=15cm}
\end{center}
\caption{Left: ``True color'' \chandra\ ACIS-I image of
CXOM31 J004327.7+411829. This image was constructed from the soft (red: 0.5--1 keV),
medium (green: 1--2 keV) and hard (blue: 2--7 keV) energy bands. The
pixel size is $0.496\arcsec$ and the image has been smoothed with a
$0.496\arcsec$ FWHM Gaussian function. Right: Hardness-ratio (1--2 keV over
0.5--1 keV) image of the
SNR, with X-ray contours overlaid. The contour levels are arbitrarily
chosen (from 0.5--7 keV smoothed image) to best follow the surface
brightness variations. Note the ``hot spots'' in the north-western arc
and southern region of the SNR.}
\end{figure*}

We searched for spectral variability in all of the sources using a
method analogous to that described in \cite*{akong-E3:pri93}, but replacing the flux
with hardness ratio. Only 12  sources show spectral variability,
corresponding to 6\% of the total population.  This is of
course a lower limit, as small changes in the spectra of weak sources
are undetectable with the number of counts accumulated in our 40~ks of
merged data. In order to further investigate the nature of the spectral variations,
we fit simple spectra to two of the brighter spectral variables.  The
fits show that as the counting rate increases, the spectrum becomes
harder (see Figure~4).  This is reminiscent of atoll and Z sources in
our Galaxy (see e.g. \cite{akong-E3:has89}). The luminosity of
these two sources ranges from $(0.4-1.0)\times 10^{38}$ erg s$^{-1}$,
which is higher than the typical luminosity ($< 10^{37}$ erg s$^{-1}$)
of atoll sources.  However, this luminosity is similar to that of the
Z~sources, which are believed to reach the Eddington limit
(\cite{akong-E3:psa95}).  The
luminosity and spectral changes appear consistent with that seen in
Z sources as they move along the ``normal branch''.  These are
probably the first extragalactic Z sources to be identified, except
for \object{LMC\,X--2} (\cite{akong-E3:sma00}). Continued
monitoring may confirm the nature of these sources by allowing us to
trace out the full Z-shape of the spectral variations expected if this
analogy is correct. Spectral variation is also seen in transients
during their evolution away from the outburst. Figure~4 shows the
spectral evolution of the brightest transient ($\sim 3\times10^{38}$ \lum)
ever found in M31 (\cite{akong-E3:kong01}); the source becomes harder
during the decay of the outburst. While the source is in its decay
stage, it is still relatively bright in a recent observations taken in
2002 January ($L_X$ \gax \, $10^{37}$ \lum).

\section{Discovery of an X-ray Resolved SNR}

We have identified two SNRs in the central region of M31 with
\chandra, and both of them were detected by \rosat\ previously. One of
them (CXOM31 J004327.7+411829) is found to be extended in X-rays as a
distinct ring-shaped object with a diameter of $\sim 11\arcsec$ (33
pc). Figure~5 shows the ``true color'' X-ray image of the SNR; also
shown is the hardness ratio image of the
SNR. It is clear that bright spots can be seen in north-western and
southern regions of the SNR. This SNR is also detected in several
optical surveys (e.g. \cite{akong-E3:dod80}; \cite{akong-E3:bla81});
it was identified as an irregular, faint and high [S\,II]/H$_{\alpha}$
($\sim 1$) SNR. The optical extension is about $10\arcsec - 18\arcsec$,
with a crescent open in the south-eastern part of the SNR. Our X-ray
image also shows similar morphology.

Initial examination of a 40ks ACIS observations on 2001 October
indicates that a non-equilibrium ionization (NEI) model with interstellar
absorption is appropriate for modeling the SNR's spectrum. It is also
evident that the spectrum is dominated by a broad emission of O VIII
line at 0.654 keV and a blend of Fe L shell lines and Ne K shell lines
around 0.9 keV. The best fitting NEI model indicates a 0.3--7 keV
emitted luminosity of $\sim 5\times 10^{36}$ \lum.

\section{Luminosity Function}

In Figure~6 we plot the cumulative luminosity function (CLF) for all
detected sources in the stacked image, and also separately the LFs of
inner bulge (region~1; central $2\arcmin \times 2\arcmin$), outer
bulge (region~2; central $8\arcmin \times 8\arcmin$ excluding
region~1), disk (region~3; central $17\arcmin \times 17\arcmin$
excluding regions~1+2), and
bulge (regions~1+2 combined).
The LF for all sources has a break at
$\sim 2\times10^{37}$ \lum, with $\alpha_1=0.38$ before
the break and $\alpha_2=1.63$ after the break. This result is in
good agreement with previous \rosat\ (\cite{akong-E3:pri93}) and \xmm\
(\cite{akong-E3:shi01}) measurements of the LF. 

The CLF for the inner bulge is significantly different, with
a break at a lower luminosity ($\sim 1.6\times10^{36}$) and a
significantly flatter distribution at the faint end. We performed a
two-sample Kolmogorov-Smirnov (K-S) test for the LFs
of inner bulge and disk, and find that there is only 3\% probability that
they are drawn from the same distribution. Simulations show that the
flattening of the CLF of inner bulge is not due to
incompleteness. 

However, the counts in a CLF are not independent and therefore direct
fits to it will underestimate the errors and may produce a biased
best estimate of the slope.  In order to more accurately estimate the
errors and slopes, we used a maximum likelihood method
(e.g. \cite{akong-E3:cra70}) to determine the slopes in the differential
luminosity functions (DLFs).  For the inner
bulge, the LF is roughly consistent with a single
power-law with $\alpha\sim0.7$ although the appearance of the LFs (Figure~6)
and the results of simulations 
indicate the presence of a brake to a flatter LF at the lowest
fluxes. However, we note that the statistics in
the inner bulge region are insufficient to confirm the presence and
constrain the size of the break.The slope of the outer bulge and disk below the break is
roughly consistent with 0.5--0.6. Further away from the inner bulge, a
break appears near $10^{37}$ \lum\ and it shifts to higher luminosity
when considering the disk region (see Figure~6). In addition, the slope becomes
steeper (from $\sim$ 1.0 to 2.2) when moving from the bulge to disk.

\begin{figure}
\begin{center}
\psfig{file=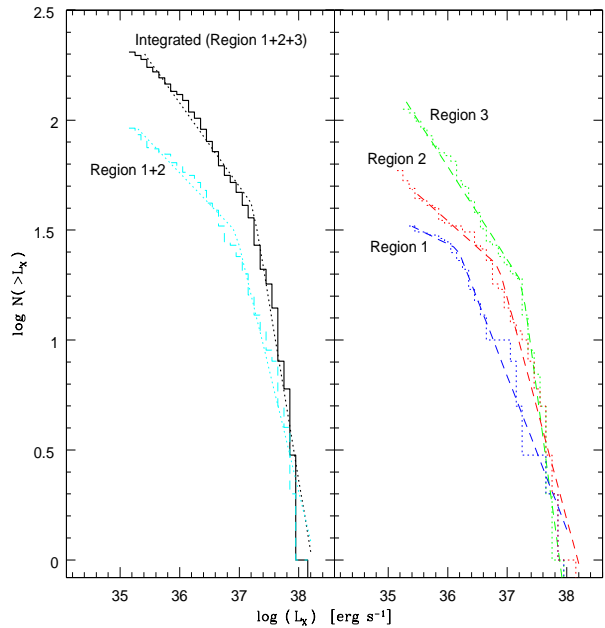,height=9cm,width=8.5cm}
\end{center}
\caption{Left: Luminosity functions for all sources (Regions 1+2+3)
and bulge (Regions 1+2).
Right: Luminosity functions for inner bulge (Region 1), outer bulge (Region 2) and
disk (Region 3).}
\end{figure}

The black holes and neutron stars that power many of the M31 X-ray
sources have formed through the evolution of initially massive stars.
Because of this, the X-ray LF traces the history of star formation and
evolution of these massive stars in binary systems.  Breaks in the LF
may indicate indicate an impulsive star formation event.  As the X-ray
binaries age, their average luminosity shifts to lower values and
therefore the location of the break may be an indication of the how
long ago the star formation event occurred (\cite{akong-E3:wu02}).
Luminosity functions which do not show a break may indicate that star
formation is still occurring.  Chandra observations have measured the
breaks in the LFs of several nearby galaxies, e.g. M81 (\cite{akong-E3:ten01}), \object{NGC\,1553} (\cite{akong-E3:bla01}), \object{NGC\,4697}
(\cite{akong-E3:sar01}) and M83 (\cite{akong-E3:sor02}), and these breaks have been
interpreted as evidence for impulsive star formation. 
Within M31, we find that the LF of three regions we studied have
breaks at a different luminosities.  The inner bulge has this break at
the lowest luminosity, and the luminosity of the break increases
monotonically as we go out from the inner bulge.  If the breaks do
indicate the epochs of star formation events, then these events
occurred most recently in the disk of M31 and further back in time as
we move towards the nucleus of M31. 

As well as the monotonic shift in the break luminosity, there is a
monotonic shift in the slopes of the LFs.  As we move in towards the
nucleus these slopes become progressively flatter.  This is 
somewhat difficult to understand in the context of the discussion
above, because the most luminous sources would be expected to have the
shortest lifetimes.  Loss of these sources as they age would tend to
steepen the luminosity function, but we find flatter
luminosity functions in the apparently older populations.  

It is interesting to compare the LF of these three regions of M31 to
those of other galaxies.  M31 is not the first galaxy found to  show a
break in its LF nor is it the first to show different LFs in different
regions; both M81 (\cite{akong-E3:ten01}) and M83 (\cite{akong-E3:sor02})
show similar behavior.  
In cases where a single slope is fit to the LF (ie, there is no clear
break) the LFs of early type galaxies
and the bulges of spirals tend to be steeper ($\alpha \sim 1.7$) than
those of spiral disks and galaxies with active star forming regions
($\alpha \sim 0.8$, e.g, \cite{akong-E3:pre01}).  
The opposite seems to be the case within M31:  the disk has a steeper
LF than the bulge region.  
We speculate that this difference may be
related to the location of the breaks in the M31 LF, which are at a
lower luminosity that those seen in other galaxies.  At these lower
luminosities we may be sampling a different class of source, and the
steepness of the LF may be due inclusion of this new class of faint
sources rather than a loss of bright sources.  We note that there is
some evidence that we are sampling a different class of sources as we
move out from the bulge, in that the fraction of sources which show
variability decreases monotonically.  If these sources have an
intrinsically steeper LF than bright accreting binaries they may be
able to cause the steepening of the LFs as we move from the bulge
towards the disk.

\section{Conclusions}

In the past two years, \chandra\ and \xmm\ have produced fruitful
results on M31 (there are 7 papers related to M31 in this
meeting). Clearly, it is just a beginning while data are still
being analyzed and will be coming in in near future. A series of
papers about the point source properties, the diffuse emission, the
X-ray emission from the central supermassive black hole, and supersoft
source populations are being prepared by our \chandra\ team. On the
other hand, an extensive \xmm\ survey on M31 is undergoing and the
results would allow us for a detailed studies of point source
properties and populations on the whole galaxy.    

\begin{acknowledgements}
We are grateful to Kinwah Wu and Andrea Prestwich for stimulating discussion and
comments. We acknowledge the support of a Croucher Fellowship (AKHK),
NASA LTSA Grant NAG5-10889 (MRG,AKHK) and NAG5-10705 (RD,AKHK), and
NASA Contract NAS8-39073 (MRG). The HRC GTO program is supported by NASA Contract NAS-38248.
  
\end{acknowledgements}

\end{document}